# Determinants of Patent Survival in Emerging Economies: Evidence from Residential Patents in India

Mohd Shadab Danish[1], Pritam Ranjan[2], and Ruchi Sharma[1]

[1]Indian Institute of Technology, Indore

[2]Indian Institute of Management, Indore

*Abstract*

The purpose of this paper is to use patent level characteristics to estimate the survival of resident patents (filed at the Indian Patent Office (IPO) and assigned to firms' in India). This study uses the renewal information of firm-level patents applied during 1st January 1995 and 31st December 2005, which were eventually granted. The data provided by IPO consists of 2025 resident patents assigned to 266 firms (foreign subsidiary firms and domestic firms). The survival analysis is carried out via Kaplan-Meier estimation and Cox proportional hazard regression. The outcomes of this study suggest that the survival length of patents significantly depends on their technological scope and inventor size. Moreover, the patents of the firms taking tax credit benefits exhibit lower survival rate as compared to patents of remaining firms. The study also finds that the patents filed by foreign firms with DSIR affiliation are getting more benefit from the R&D tax incentive policy.

JEL: 031; 034

Contact: MOHD SHADAB DANISH*, PRITAM RANJAN†, and RUCHI SHARMA‡,
*Assistant Professor in Economics, Dr BR Ambedkar School of Economics University, Bengaluru, India (Email: shadab@base.ac.in), †Professor in OM&QT, Indian Institute of Management Indore, India (Email: pritamr@iimidr.ac.in), ‡Professor in Economics, Indian Institute of Technology Indore, India (Email: ruchi@iiti.ac.in)

## 1. Introduction

Anyone using patent data to measure the output of innovation faces two fundamental challenges. First, all patents are not equally important or valuable and therefore measuring innovation output using raw patent count is misleading. Second, due to non-availability of any formal market for the sell and purchase of patents, it is difficult to assign the actual value to the patents. The literature on patent value has proposed various indicators that are correlated with the value of patents (Trajtenberg 1990; Hall, Jaffe, and Trajtenberg 2005). The high-value patent in this study is considered to have a longer active life (see Schankerman and Pakes, 1984). Patent lives are divided into three phases: first, starts from the date of filing to examination date; second, from the date of request to the grant date; and third, is the period between grant dates to lapse date, that is, the total active life of the patent (Nikzad, 2011). Whereas, Masurseth (2005) and Sevenseeon (2007) divided patent life into two phases: pre-grant period called 'provisional' life, and the post-grant period called 'active' life. In order to protect the entitlement of patent, patent offices' charges annual renewal fee from the inventor. The renewal fee is one of the significant sources of income for IPO. Patent life depends upon the value generated during the protection period for inventor against compulsory renewal fee. Hence valuable patents are more likely to be renewed. Schankerman and Pakes (1985) argued that the renewal decision of patent is purely based on economic criterion. In many countries including India, patent holders are required to pay an annual renewal fee to keep their patents active, which makes sense only if the renewal cost is lower than the value generated by those patents. Also, if the patent loses its market relevance due to more advanced and valuable invention, then it allowed lapsing by the inventors. Thus, the renewal information reveals the quality/value of the patent in the country. The survival length of the patent shows essential information regarding firms' R&D quality, log-term advancement and strategy towards intellectual property. Note that the weak quality patent has shorter survival length and hence considered to have a lower value (Schankerman and Pakes, 1984).

An important question with this respect is whether high survival length across the technology group has any common identifiable characteristics such as higher patent claims, family size or inventor size. Zeebroeck and Pottelsberghe (2007) conduct exploratory research on the survival length of the patent where they find that more valuable patent is more cited and covering a larger geographical scope. However, a higher number of patent claims is associated with longer grant lag also with a lower chance of grant and lower renewal rates. Svensson (2012) suggested that high-quality patents (measured by a higher number of citations, litigated patents, boarder technology scope) would have a higher probability of being both renewed and commercialized. The renewal length of the patents typically depends on the quality of the R&D, marketability of the invented product, license or sale of the invention, and nature of the technology (see Pakes and Simpson, 1989; Tong and Frame, 1994). In the recent past government came up with various policies to promote innovation as well as its quality. However, there is no study to the best of our knowledge discussing the role of R&D tax credit policy on the patent life from an Indian perspective.

The objective of this study is to estimate the determinants of survival length of a patent granted by IPO. This study contributes to survival literature in two different aspects. First, this study analyses various driving factors of survival length across the different technological group. Second, it includes R&D tax credit policy as one of the determinants of the survival length. The tax credit on R&D investment provides financial benefits to the firms to focus on cutting-edge technology. Therefore such patents are assumed to be less vulnerable relative to non-DSIR affiliated firms' patents, and hence should lead to longer patent life. Thus, based on the literature, the determinants of patent survival are clustered into four groups. First, the *complexity of the inventions* is measured by patent technology class (4-digit IPC class), number of inventors, and the grant lag. Second, the *filing strategy* includes the structure and quality of the drafted document (number of claims) and protecting the same patent in a different jurisdiction (family size). Third, *ownership characteristic* that is firms' country affiliation is included in the study. Fourth, *R&D tax credit policy* of the government given to eligible firms on in-house R&D investment.

There is several methodologies proposed to analyze survival data. The Kaplan-Meier curves and Cox proportional hazards (Cox-PH) regression is popular among others used frequently by scholars across different application areas. Kaplan-Meier curve estimation is a non-parametric statistical tool whereas Cox-PH model is a semi-parametric model. This study focuses on the survival function estimates for patents based on their affiliation with DSIR vs non-DSIR firms. The Cox-PH model is applied to get an in-depth impact analysis of patent characteristics on the survival length. The number of patent and ownership characteristics such as number of claims (NC), number of inventors (NI), family size (FS), technology scope (TS), DSIR dummy, and ownership dummy (OW) is applied in the Cox-PH model.

The econometric model is used on the data set of all granted patents (resident) that were filed at IPO between 1st January 1995 and 31st December 2005. We have a sample of 2025 residents out of 4343 patents assigned to 266 firms (domestic and foreign subsidiary) with complete information. The patents are classified into five categories (electrical, mechanical, instruments, chemistry and other fields) based on international patent classification (IPC). Applying both Kaplan-Meier and the Cox proportional hazard model to the data, this study finds that overall non-DSIR firms' patents are more likely to survive for a longer time. As per the hypothesis patents of DSIR, affiliated firms survive longer; however, the results suggest otherwise. The finding of ownership category shows that patents of foreign affiliates in India are less likely to survive yet the interaction between ownership and DSIR (ownership*DSIR) in our Cox-PH model finds a positive impact on the survival length of the patent. Thus, the overall domestic firms produce low-quality patent in-comparison to foreign subsidiaries in India. Electrical and mechanical patents are more likely to survive as compared to chemistry and instruments. The impact of technology class on survival length suggests that if a patent belongs to more than one technology class, it is more likely to survive longer. Whereas, geographical scope (family size) and drafting style (number of claims) have no significant impact on the firms' patent survival rate.

The rest of the manuscript is organized as follows. Section 2 presents an overview of the study on the determinants of patent survival and brings the necessary discussion on the contradictory results reported by various studies. Section 3 presents the data and descriptive statistics. In Section 4, we propose the empirical model. The estimated results are presented in Section 5. We conclude our results in Section 6.

## 2. Literature

A brief review of different literature related to survival analysis from the valuation perspective is presented in section 2.1. The survival studies find that a longer patent life is an indicator of higher value patent. Section 2.2 presents the number of quality an indicators that influence the survival length of the patents. Section 2.3 briefly discusses the R&D tax credit policy in India's context.

## 2.1. Patent survival

Patents count considered to be the weak proxy of innovation (Trajtenberg, 1990). The simple patent count does not consider the heterogeneity among the patents. Therefore, we often end up making the wrong judgment about the quality and value of innovation. However, disaggregated information revealed in the patent documents brings richness in the patent data. Over the period renewal length of a patent is studied by many scholars to estimate the value of patents (see Pakes and Schankerman, 1984; Schankerman, 1998, Lanjouw et al., 1998). To keep a patent alive after issuance, the patentee must pay the renewal fee. The renewal fee varies with the age of patent and the corresponding patent offices. In return, patent generates implicit profit to the patent owner during the coming year. However, if the patent renewal fee is not paid, the patent expires permanently and therefore, after the return on that patent becomes zero (Lanjouw et al., 1998).

Most of the previous studies have used patent renewal information to estimate the value distribution of patents (Schankerman and Pakes, 1986; Griliches, 1990; Bessen, 2008). The literature on patent valuation finds that patents who survive longer have a higher value compared to patents lapsed in the early age (Bessen, 2008). It is assumed that owners' are well aware of the usability and quality of the patent, and the decision about the renewal of patent is based on the economic principle (Svensson, 2012). The patent renewal decisions of owners' are influenced by many other uncontrolled factors such as future marketability of the patented products, and the advancement over the earlier invention and so on. Schankerman and Pakes (1986) estimate the distribution of patent value where they find that about half of the European patents continue to be renewed after ten years and only 10 per cent of patent lives up to statuary term. Griliches (1990) finds that the patent with lower value depreciates rapidly, and only a few patents qualify to high value.

There are other strands of literature that focus on the determinants of patent renewal. Harhoff et al. (1999) find that patents renewed to statuary period are cited more often than patent ceased/expired in early age. Serrano (2008) finds that the acquired patent is more likely to be renewed than non-acquired ones. Maurseth (2005) uses a survival model to estimate the determinants of patent renewal. The result suggests that patent cited across the technological field survive longer than patent cited within the technological field.

## 2.2 Determinants of patent survival

The selection of the explanatory variables and the sampling methodology varies widely across the studies. To start with some of the explanatory variables- patent citation, the number of claims, family size, and technology scope- that is proven reliable when patent value distribution or patent survival measured (value and survival is interchangeably used). The valuation studies have extensively used citation information along with legal disputes and renewal information to measure the value of patents (Moore, 2005; Allison, 2003). The patent value determinants are grouped into four different categories of variable in the equations: (1) different characteristics of patent application (PC), (2) ownership characteristics (OC) (3) some contextual information collected through survey if any (4) and the filing strategy inventors (FS) (van Zeebroeck and van Pottelsberghe, 2011).

In many cases, explanatory variables have been used in both sides of the equation depending upon the underlying objective of the study. The forward citation counts are derived measure, which indicates about the quality and value of the patent. The patent family size identifies as a measure of geographical scope, a measure of the patent length (renewal years) and the litigation information are often used as dependent as well as an independent variable in valuation studies. Van Zeebroeck and van Pottelsberghe (2011) reviewed several studies on patent valuation that precisely used these four variables.

The backward citation (measure of existing technological background) (Reitzig, 2004), non-patent citation (basic research) (Narin et al., 1987), number of claims (legal breadth of the patent) (Tong and Frame, 1994), number of technology class (technological scope) (Lerner, 1994) and the number of inventors (indicating the research efforts) (Brusoni et al., 2006) have been frequently used as a determinants of patent value. Brusoni et al., (2006) study show that the number of inventors is strongly associated with the size of the applicant firm. This indicates a difference in the scale of research activities among the organization and firm'. Thus, inventor size is often used as a proxy of firm size in the studies. Zeebroeck et al. (2011) used some of these characteristics as a complexity indicator in their study and found them positively associated with the patent value. The presence of multiple applicants denotes the joint research efforts (Duguet and Iung, 1997), cross-border ownership indicates international collaborations (Guellec et al., 2000), are expected to associate with the patent value positively.

Gambardella et al., (2008) find that independent inventors to large multinational firms have an ambiguous relationship with patent value. On the other hand, academic patents are related to more basic research, which may have higher scientific value (Harhoff et al., 2002) but have very limited takers in the market because it has a lot of uncertainty in the market. Allison et al. (2003) proposed inexperience patentee (new to the patent system) as one of the determinants of patent value. Shane (2011) uses the firms' patent portfolio as an indicator of the level of experience with the patent system as a determinant of patent value. The result suggests that the patent portfolio is positively related to the value of patents. The high patent portfolio also indicates the higher propensity to patent, possibly encouraging many patent filing of a lower value.

The present study is based on an event (survival duration) where survival analysis (Cox proportional hazard model) is used to estimate the function (Cox, 1972). The event refers to the year in which a patent expires. The censored patents are those patents which reach to 20 years or has not expired during the study period. Svensson (2011) studies the impact of the different explanatory variable (e.g. commercialization decision, patent quality, firm size, etc.) on the patent length. The study surveys Swedish patent granted to firms and individuals in 1998. The result shows that commercialization and defensive strategies increase the probability that patent to be renewed. Svensson (2011) study is based on the patents owned by small firms and individuals. Therefore, the results of this study cannot be generalized to all firms', country or region. However, the present study overcomes this issue by conducting a comprehensive study on both big and small firms' patents in India. To best of our knowledge, there is no literature available on the survival of Indian patents particularly using renewal information along with in-depth patent level information.

## 2.3 Tax credit policy

Government of India's department for scientific and industrial research (DSIR) provides a tax credit to the firms on the R&D investment under Section 35(2AB) of the Income-tax Act, 1961. Section 35(2AB) provides a weighted tax deduction of 150 per cent on in-house R&D[1]. The guideline specified that companies are allowed to claim a deduction of capital investments on R&D centre of more than Rs. 10 million (excluding expenditure on land and building). This initiative was started by the government in the year 2001 to attract innovation and technological advancement in its jurisdiction. This policy initiative has immensely influenced the R&D activities and witnessed some robust growth in innovation activities. Many progressive steps have been taken by the government of India to promote newer R&D investment in India by Indians as well as potential foreign investors. For example, increase in R&D support, improvement in the pool scientific manpower, providing good R&D infrastructure, Establishment of research facilities and centres of scientific excellence on par with some of the most globally renowned facilities (Worldwide R&D Incentives Reference Guide 2018)[2].

---

[1] 200% upto AY 2017-18

The stated goal of tax credit policy is to encourage local research and development (R&D), to incentivize businesses through IP, and to reduce the cost of R&D. Since we know that R&D funding is one of the crucial factors that influence innovation across the sectors, tax credit policy on R&D brings relief. Thus, we hypothesize that affiliation with the DSIR positively influences the quality of the patent. Hence the survival of the patents is higher. Jose et al., (2019) study found that DSIR affiliation improves the R&D and patenting activities in India. However, the impact of DSIR affiliation on the quality and value of the patent has not been investigated yet. This study divides patents into two categories one owned DSIR companies and other non-DSIR companies. We want to see how the survival length of the patent varies between these two categories.

## 3. Data and Variable Description

The data used in this study consist of all firm-level resident patents (assignee country India) applied between 1st January 1995 and 31st December 2005 that was eventually granted by the Indian patent office (IPO). The total number of patents as per IPO is 4343. The granting period of these patents was between 1st January 1997 (minimum two years since the filing date) and 31st December 2018 (the data collection date). The patent level information was collected from the IPO website[3] and PatSeer[4]. Table 1 summarizes the patent level characteristics for this data. We only considered the patents with complete details on renewal length and these patent characteristics. Note that the restriction on the completeness of the data further reduced our sample size to 2025. The dataset represents 266 firms and five technology groups (Chemistry, Electrical, Mechanical, Instruments and Others).

---

[2] https://www.ey.com/Publication/vwLUAssets/ey-worldwide-rd-incentives-reference-guide-2018/$FILE/ey-worldwide-rd-incentives-reference-guide-2018.pdf
[3] https://ipindiaservices.gov.in/publicsearch
[4] A private data base company license owned by IIT Indore.

Several patent characteristics have also been discussed in the literature earlier, for instance, Xie and Giles (2011), and Zeebroeck and Pottelsberghe de la Potterie (2011). However, we have included additional features like ownership characteristics and grouped the patients into five different technology categories based on 4-digit IPC classification.

The dependent variable in this study is the patent renewal duration (or survival length). From a survival analysis standpoint, if a patent has expired, then it coded as 1 (and referred to as the occurrence of the "event"), whereas if the patent has either matured (completed 20 years of renewal life) or still in force at the time of data collection (31$^{st}$ December 2018), then the patent is coded as 0 (and referred to the non-occurrence of the event).

## 4. Model Description

Survival analysis or failure time data analysis has been used in many disciplines and applications ranging from the actual lifetime of a patient (Lawless, 1982) to economic quantities (Kiefer et al., 1988, and Haurin and Sridhar, 2003), labour dispute (Kennan, 1985) and monetary policies (Yu, 2005). However, as per our knowledge, there is no survival study on the renewal length of patents with respect to different characteristics.

Here, the objective of survival analysis is to model the underlying distribution of the failure time, T, which is patent expiration event due to non-payment of renewal fee under 20 years of patent life from the date of filing. The dependent variable (renewal years) is assumed to have a continuous probability distribution $f(t)$ with

$$F(t) = Prob(T \leq t) = \int_0^t f(s)ds. \quad (1)$$

The corresponding survival function is $S(t) = 1 - F(t) = Prob(T \geq t)$, and the hazard rate (or hazard function) can be estimated via

$$h(t) = f(t)/S(t). \quad (2)$$

As per the literature, a plethora of methodologies have been proposed to analyze survival data, and, Kaplan-Meier curves and Cox proportional hazards regression are very popular tools across different application areas.

Kaplan-Meier curve estimation is a non-parametric statistical method used for estimating the survival function. That is, S(t) can be approximated as

$$\hat{S}(t) = \prod_{\substack{j=1 \\ (t_j<t)}}^{N} \left(1 - \frac{d_j}{n_j}\right), \qquad (3)$$

where $t_j$ is the j-th event time (basically, year-end) when the patent expires, $d_j$ is the number of patents expire at $t_j$ and $n_j$ is the number of patents known to have survived up to time $t_j$. Typically, the estimated survival curves are used to assess the impact of important grouping variables. In this paper, we focus on the comparison of survival function estimates for patents based on their affiliation with DSIR versus non-DSIR firms.

For a more in-depth impact analysis of patent characteristics on the survival length, we use Cox proportional hazard regression model (in short, referred to as Cox-PH model). The model is expressed by the hazard function denoted by $h(t)$, which measures the risk of getting a patent expired at time $t$. The hazard function is characterized by a set of time-invariant covariate vector $x_i$, and a time-dependent baseline hazard $h_0$ – which corresponds to the value of hazard rate if $x_i$ is equal to zero. The model is written as follows:

$h(t|x_i) = h_0(t) \cdot \exp(x_i^T \boldsymbol{b}) = h_0(t) * \exp(b_1 \cdot NC + b_2 \cdot NI + b_3 \cdot FS + b_4 \cdot TS + b_5 DSIR + b_6 \cdot OW + b_7 \cdot D_{chem} + b_8 \cdot D_{elec} + b_9 \cdot D_{mech} + b_{10} \cdot D_{inst} + b_{11} \cdot D_{other} + b_{12} \cdot OW * DSIR),$ (4)

where $x_i$ is a 12-tuple vector of covariate values that correspond to the patent characteristics: NC, NI, FS, TS, DSIR and OW, described in Table 1, and $D_{chem}, D_{elec}, D_{mech}, D_{inst}$ and $D_{other}$ – the dummies for the five technology categories as per the 4-digit IPC classification.

A popular summary statistic of interest, called the hazard ratio, is defined by $\exp(b_i)$, which implies that if the $i^{th}$ covariate value increases, the hazard (or the chance of patent expiring) increases and thus, the length of survival decreases. More precisely if the value of hazard ratio is greater than one, the covariate is positively associated with the event probability and negatively associated with the length of survival.

One can use a variety of approaches to assess the validity of the proportionality assumption of Cox-PH model, for instance, the graphical techniques based on Schoenfeld residuals, and tests built using hazard ratios. Even the Kaplan-Meier curves can indicate the violation of the proportionality assumption. See Harrell and Lee (1986) for details on the premise of Cox-PH model.

## 5. Results Discussion

In this section, we first present the Kaplan-Meier curve analysis of patent survival length data concerning DSIR affiliation and then discuss the impact analysis of different patent characteristics on survival length via Cox-PH model.

### 5.1 Kaplan-Meier Approach

Recall that the paper makes a patent policy level hypothesis: the patents filed by firms that are DSIR affiliated are more likely to be worthy as compared to firms not affiliated with DSIR, and it is intuitive to assume that a worthy patent has a longer survival/renewal life. We use all 2025 firm-level resident patent data obtained from IPO filed between 1$^{st}$ January 1995 and 31$^{st}$ December 2005 (and eventually granted) to estimate the Kaplan-Meier survival curves under the two categories (DSIR and non-DSIR). Figure 1 shows the estimated survival function of all patents. The plots are drawn using STATA software.

Figure 1. survival curves obtained through Kaplan-Meier estimator

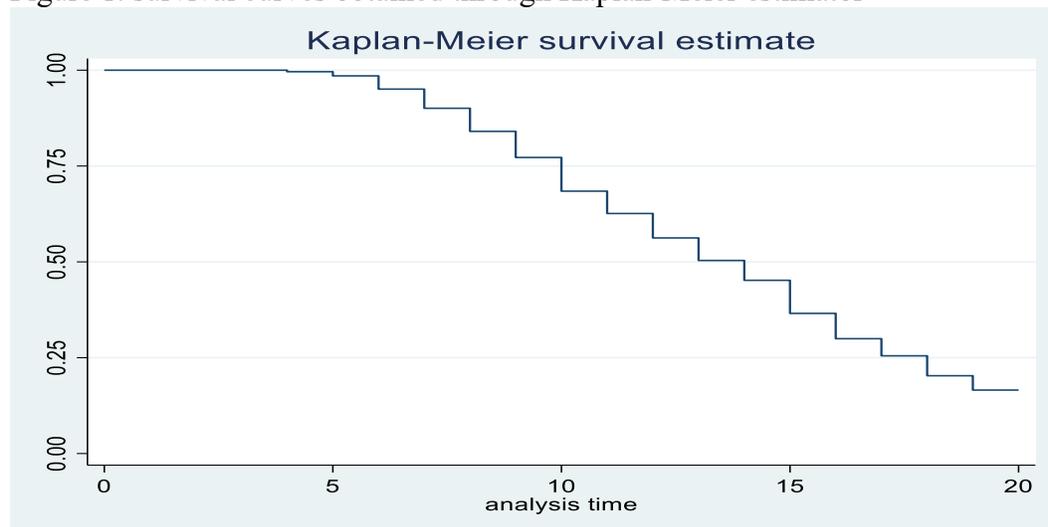

The patent belonging to DSIR firms expire at a slightly faster rate as compared to non-DSIR firms. Also, the divergence between DSIR and non-DSIR firms patents are smaller at the beginning, which increases subsequently. This implies that after the 12$^{th}$ year survival rate of the DSIR patents decreases faster. One possible interpretation could be that such firms file a lot of patents, but many of them do not have very high quality, and hence tend to lapse a little early. Also, big firms which involved in the higher R&D activities are possible to end up increasing patent portfolio to attract investment through an initial public offering (IPO) then creating a quality patent. However, for the deeper understanding of DSIR impact on survival length in the Cox-PH model, we interacted ownership with DSIR dummy. The result is presented below in Table 2.

We also conducted "log-rank test" to test the difference between DSIR ($n = 148$) and non-DSIR firms ($n = 118$) (Harrington, 2005). The results obtained show a statistically significant difference between the survival rates for the two groups of patents $\chi^2(1) = 23.38, p \approx 0.00$.

### 5.2 Cox Proportional Hazard Model

The main objective of this section is to assess the impact of the following patent characteristics, number of claims (NC), number of inventors (NI), family size (FS), technology scope (TS), DSIR affiliation, ownership (OW), and the dummies for the five technology categories (Chemistry, Electrical, Mechanical, Instruments and 'Other field') as per the 4-digit IPC classification. The dependent variable represents the probability of a patent expiring at the time (t). Table 2 summarizes the model fit under different scenarios. Model 1 corresponds to a simpler case with only DSIR as the independent variable, and Model 2 – Model 6 include all additional five patent characteristics (NC, NI, FS, TS and OW), but focus on only one technology at-a-time. Finally, Model 7 assumes all technologies (with Instruments as the reference category) and all patent characteristics.

Table 2. Cox proportional hazard regression model

| Variable | Model 1 Coef. | Model 2 Coef. | Model 3 Coef. | Model 4 Coef. | Model 5 Coef. | Model 6 Coef. | Model 7 Coef. |
|---|---|---|---|---|---|---|---|
| DSIR | 0.330*** (0.07) | 0.310*** (0.08) | 0.321*** (0.08) | 0.266*** (0.08) | 0.290*** (0.08) | 0.318*** (0.08) | 0.305*** (0.08) |
| Claims (NC) |  | 0.003 (0.00) | 0.003 (0.00) | 0.001 (0.00) | 0.002 (0.00) | 0.003 (0.00) | 0.001 (0.00) |
| Inventor size (NC) |  | 0.026** (0.01) | 0.026** (0.01) | 0.018 (0.01) | 0.021 (0.01) | 0.025*** (0.01) | 0.017 (0.01) |
| Family size (FS) |  | 0.001 (0.00) | 0.002 (0.00) | 0.000 (0.00) | 0.001 (0.00) | 0.002 (0.00) | 0.000 (0.00) |
| Technology scope (TS) |  | -0.009** (0.00) | -0.009* (0.00) | -0.013*** (0.01) | -0.011** (0.01) | -0.009* (0.00) | -0.014*** (0.01) |
| Ownership (OW) |  | 0.036 (0.14) | -0.015 (0.14) | 0.039 (0.14) | -0.036 (0.14) | -0.013 (0.14) | 0.344** (0.19) |
| Electrical |  | -0.228** (0.11) |  |  |  |  | -0.348*** (0.16) |
| Instrument |  |  | 0.096 (0.13) |  |  |  |  |
| Chemistry |  |  |  | 0.306*** (0.06) |  |  | 0.044 (0.13) |
| Mechanical |  |  |  |  | -0.287*** (0.07) |  | -0.294** (0.14) |
| Other field |  |  |  |  |  | -0.198 (0.18) | -0.263 (0.22) |
| Ownership*DSIR |  |  |  |  |  |  | -0.633** (0.29) |
| LR $\chi^2(1)$ | 22.24*** | 34.9*** | 30.73*** | 56.38*** | 48.14*** | 31.42*** | 39.56 |
| No. of observation | 2,025 | 2,025 | 2,025 | 2,025 | 2,025 | 2,025 | 2,025 |
| PH assumption $\chi^2(7)$ | 0.22 | 9.10 | 7.45 | 12.05 | 3.70 | 3.07 | 20.06 |

**Notes:** Standard errors are shown in parentheses. Parameters are significant at, *** $p < 0.001$, **$p < 0.05$, *$p < 0.01$

A few quick remarks as per Table 2 are as follows:

1. The coefficient of DSIR is consistently positive and statistically significant, which implies that DSIR affiliated firm patents (coded as 1) as compared to non-DSIR firm patents (coded as 0) have a greater hazard rate and hence lower survival rate. This inference is consistent with the Kaplan-Meier curve estimate comparison.

2. The following patent characteristics: number of claims (NC), family size (FS) and technology scope (TS), have the same sign of parameter estimates and are statistically significant across different models, except ownership (OW) and inventor size (NI).

3. The coefficient of inventor size is positive and significant, which implies that the larger the inventor size, the higher the hazard rate is. This appears to be counter-intuitive as compared to some other studies, for example, Brusoni et al. (2006).

4. In the technological category, electrical and mechanical patents are more likely to survive for a longer time compared to chemistry, instruments and other field. This result corroborates with Danish et al. (2019) study on Indian patent valuation.
5. The negative coefficient of the interaction term between ownership and DSIR (in model 7) shows that foreign subsidiary firms with DSIR affiliation have a positive impact on the survival length as compared to Indian firms with DSIR affiliation.

Overall result transpires that some patents owned by firms are having a longer life and some patents lapses early. Therefore, while measuring the innovation effect on the various economic indicators should not be weighted equally.

## 6. Conclusion

The main ingenuity of this study is to investigate the determinants of patent survival in a systematic way by utilizing a rich set of data on Indian patents assigned to both domestic and MNCs from 1995 to 2005. The first outcome of this study reveals that an increase in the average patent length is not only influenced by the patent characteristics, but the technological and ownership characteristics are equally important. The delays in the granting process of the patent not necessarily associated with the higher renewal rate.

Other finding reveals that patents owned by DSIR affiliated firms are more likely to lapse early and therefore, the survival rate is lower, and the hazard rate is found to be higher. Although tax credit policy (DSIR affiliation) is beneficial for the firms, the life of the patent is not improved relative to non-DSIR firms. Geographical scope (family size) and drafting style (number of claims) have no significant impact on the firms' patent survival rate. However, if a patent belongs to more than one IPC (4-digit) technology group, the survival rate is found to be higher, and the hazard ratio lower. This implies that broader technological patents in India are more likely to survive. In technological categories, the result suggests that electrical and mechanical patents have a lower hazard rate and higher survival rate relative to instrument patents.

In contrast, chemistry and 'other field' patents found insignificant. The longer survival of a patent can be understood as a high-value patent. Foreign subsidiary firms in India benefit more from the government tax credit policy. The result suggests that DSIR affiliated foreign subsidiary firms' patent hazard ratio is smaller compared to non-DSIR firms and therefore, the higher survival rate is observed for such patents. One could also investigate other flexible models like survival trees and survival random forest. The future work can be extended to all patents (residential and non-residential) in India.